\documentclass{ametsocV6.1}
\usepackage{xcolor}
\usepackage{amsmath}

\title{Blended parameterization in an atmospheric model: Improving severe storm ensemble prediction by considering uncertainties in model physics}

\authors{Khanh Hung Mai\aff{a}\correspondingauthor{Khanh Hung Mai, maikhanhhung18988@gmail.com}
Duc Le,\aff{b} 
Saito Kazuo,\aff{c} 
Futo Tomizawa, \aff{b} 
Yohei Sawada,\aff{b} 
}

\affiliation{\aff{a}{National Center for Hydro-Meteorological Forecasting, Vietnam}\\
\aff{b}{Department of Civil Engineering, University of
Tokyo, Tokyo, Japan}\\
\aff{c}{Atmosphere and Ocean Research Institute, University of Tokyo, Chiba, Japan}\\
}

\abstract{
{Physics parameterizations are often needed for numerical weather prediction (NWP) of precipitation forecast. This is mainly because the resolutions of most computational atmospheric models are not fine enough to explicitly resolve sub-grid scale processes associated with precipitation systems. The various options in each physical parameterization scheme introduce model physics uncertainty, leading to variations in simulated precipitation due to differing representations of physical processes. We aim to quantify and reduce uncertainties in severe storm prediction arising from selecting physics parameterization schemes. In this study, we introduced a method called ’blended parameterization’ in an atmospheric model. This method parameterizes the selection of physical parameterization schemes using weighting parameters. This approach reduces the model selection problem to the optimization of these weighting parameters.}
The Markov Chain Monte Carlo (MCMC) method was used to estimate the posterior probability distribution of the weighting parameters. The large computational cost of MCMC was resolved by a surrogate model that efficiently mimics the relationship between weighting parameters and likelihood. Subsequently, the weighting parameters were sampled from the posterior distribution, followed by conducting an ensemble simulation to predict rainfall in Vietnam. Our optimized “blended parameterized” ensemble prediction system outperformed a conventional physical ensemble prediction system similar to that operationally used in Vietnam.}

\begin{document}
\raggedbottom
\maketitle

%
%
%
\statement
{
Our research presented a novel approach to address the limitations of the numerical weather prediction (NWP) system in assessing and reducing uncertainty in selecting the physical parameterization scheme. This study applied the blended parameterization approach to an atmospheric model for the first time to improve severe rainfall forecasting. We successfully optimized the weighting parameters of the blended parameterization with a reasonably small computational cost using a surrogate model. Our results show that the newly developed blended parameterization-based Weather Research and Forecasting (WRF) model outperformed the operational ensemble prediction system of Vietnam National Center for Hydro-Meteorological Forecasting (NCHMF). 
}

%
%
%

%
\section{Introduction} 
\label{sec:Introduction}
{
Precipitation prediction is a crucial component of weather forecasting \citep{croitoru2013} and often relies on numerical weather prediction (NWP) models. Despite extensive research and significant efforts to develop sophisticated NWP models, achieving accurate rainfall forecasts remains a major and persistent challenge. This difficulty arises from the inherently complex and nonlinear behavior of the atmosphere. Additionally, many physical processes that govern precipitation, such as cloud formation and development, occur at sub-grid scales smaller than the resolution that most current computational atmospheric models can explicitly represent \citep{wang2021quantifying}. This requires the use of parameterizations to account for these sub-grid scale processes.
} 

{
Rainfall prediction using NWP models is influenced by three main uncertainty sources, including uncertainties in the input data which encompass initial conditions, uncertainties in model parameters referring to the physical constants used within the model and uncertainties in the structure of the models, relating to the representation of physical processes within the NWP model \citep{moosavi2018,XuanWang2023}. The uncertainty in initial condition amplifies over time due to the chaotic nature of the atmosphere \citep{lorenz1963,lorenz1965}. Conventional data assimilation methods, which aim to bring the initial state of the NWP model closer to the available observations can address this uncertainty and have proven effective in precipitation forecasts \citep{zhang2006,li2020quantitative}. Another significant source of uncertainty arises from model parameters, which are often physical constants that cannot be directly observed \citep{wang2023}. To address this, automatic parameter optimization methods have been applied to NWP models \citep{afshar2020uncertainty,chinta2020calibration,chinta2021assessment,di2015assessing,di2017parametric,di2018assessing,Di2019}.
} 

{Beyond the uncertainties in initial conditions and model parameters, a critical source of forecast uncertainty stems from the model physics, specifically selecting the appropriate physical parameterization schemes \citep{mcmillan2012benchmarking,gupta2012towards,evin2014comparison, moosavi2018}. 
Because there are many crucial physical processes for precipitation that occur at scales smaller than the computational grid of NWP models, they must be simulated through simplified physical representations known as physical parameterization schemes. NWP models offer numerous options for physical parameterization schemes for various physical processes, including cloud microphysics, cumulus convection, radiation, and planetary boundary layer processes. Different physical parameterization schemes represent the underlying physics in different ways, which can lead to variations and uncertainties in the resulting rainfall forecasts \citep{lowrey2008assessing}. The sensitivity of simulated rainfall to the selected physical parameterization schemes has been extensively documented in previous works \citep{lowrey2008assessing, nasrollahi2012assessing, du2019impacts}.
}

{
Despite the significant progress in developing methods such as data assimilation and parameter optimization to handle initial condition and  parametric errors, the uncertainty arising from the selecting of physical parameterization schemes in NWP models remains a considerable challenge that has not been fully addressed by previous research efforts \citep{wang2021quantifying, mai2020simultaneously}. Many earlier studies attempting to resolve this issue have heavily relied on expert judgment to select a limited number of combinations of physical parameterization schemes for evaluation, which restricts the exploration of the vast model physics uncertainty space \citep{moosavi2021machine,wang2021quantifying}. Furthermore, these investigations typically assessed uncertainty over a discrete and countable model ensemble space, where each ensemble member represented a forecast with a specific and fixed combination of physical schemes \citep{mai2020simultaneously}. This discrete evaluation approach does not provide a deep understanding or a comprehensive quantification of the model physics uncertainty \citep{wang2021quantifying}. Moreover, specific physical configurations that demonstrate good performance in particular weather situations may not be suitable or accurate for other conditions, and no single configuration has been found to accurately predict severe rainfall  across all times and locations \citep{evans2012evaluating,du2019impacts,wang2021quantifying}.
}

{
To address the aforementioned limitation of quantifying and mitigating uncertainties when selecting physical parameterization schemes of NWP models, this study introduces a novel approach based on blended parameterization. The main idea of the blended parameterization approach is to parameterize the selection physical parameterization schemes by introducing weighting parameters. Instead of choosing a single scheme for each physical process, the blended parameterization reformulates a model to simulate its physics by weighting averages of different physical parameterization schemes that are chosen simultaneously. This approach reduces the discrete model selection problem to the optimization problem of weighting parameters \citep{mai2020simultaneously}. Subsequently, Markov Chain Monte Carlo (MCMC) is employed to determine the posterior probability distribution of these weighting parameters conditioned by rainfall observation. By using the blended parameterization approach, this study aims to provide a more comprehensive quantification of the uncertainties associated with the selection of physical parameterization schemes in NWP models for improved precipitation forecasting.
}

\section{{Model, data, study area and study cases}} 
\label{sec:Model, data and study area}
\subsection{{Model}}
\label{sub:Model}
{The high-resolution mesoscale meteorological model selected for this study is the Weather Research and Forecasting Model (WRF) model, utilizing the Advanced Research WRF (ARW) dynamic core version 3.9.1. The model is run for 72 hours, with outputs generated at a horizontal resolution of 10 km, a total of 325 × 283 grid points, and 40 vertical pressure levels and outputs provided at one-hour intervals. In this study, the WRF model has been reformulated to parameterize the selection of physical parameterization schemes using weighting parameters. The physical parameterization schemes employed include microphysics, cumulus, planetary boundary layer, shortwave radiation, and longwave radiation. Details of  this reformulation will be provided in section \ref{sub:Multi-physics ensemble by blended parameterization}}

\subsection{{Data}}
\label{sub:Data} 
{\textit{a. Observation Data}}

\begin{figure}[h]
 \centerline{\includegraphics[width=25pc]{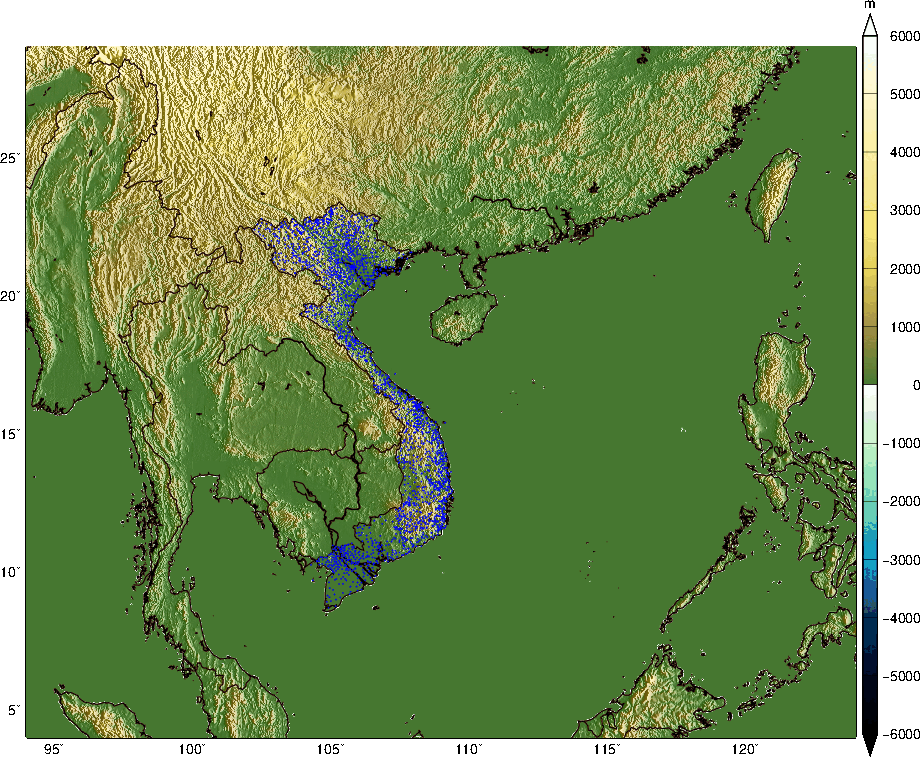}}
  \caption{Model domain and Vietnamese rain gauge station locations (blue dots)}\label{fig:domain}
\end{figure}

{
The observation data used in this study are rainfall data from approximately 2,500 rain gauge stations across Vietnam (Fig. \ref{fig:domain}). The data is interpolated to the NWP model grid points to evaluate the model's output. The time interval is one hour. The rainfall observations were used for parameter optimization and uncertainty assessment.}

{\textit{b. Initial and boundary conditions data}}

The Final Global Analysis (FNL) data comes from the Global Data Assimilation System (GDAS), which gathers observational data from the Global Telecommunications System (GTS) and other sources. FNL data are generated every three hours on a 0.25° grid \citep{cisl_rda_dsd083003}. FNL data uses the same model as the Global Forecast System (GFS) but is prepared about an hour later to include more observational data. The analyses cover various levels from the surface to the tropopause, including pressure, temperature, wind, humidity, and ozone. FNL data are widely used in meteorological studies \citep{li2020quantitative,Liu2021}. In this study, we chose the FNL data as the initial and boundary conditions for the regional high-resolution WRF model.

\subsection{{Study area}} 
\label{sub:study_area}
The forecasting domain is configured based on the causes of rainfall in Vietnam \citep{Yokoi2008,Chen2012,vanderLinden2017}. It spans from  4°–29°N and 94°E–124°E (Fig. \ref{fig:domain}). This simulation domain covers the entire Vietnam region, extending slightly further northward and eastward to include the surrounding seas, particularly the East Sea. This domain allows the model to effectively represent synoptic processes causing heavy rainfall in the Vietnam region. 

\subsection{{Study Cases }} \label{sub:study_cases}
We examined five severe rainfall events that occurred in Vietnam in 2022 examined in our study. The selection of these events was based on 72-hour accumulative rainfall data. {These events are highly representative to severe rainfall in Vietnam. The causes of precipitation for these five events are cover the main synoptic weather patterns identified in previous studies \citep{Yokoi2008,Chen2012,vanderLinden2017,du2019impacts}, including low-pressure troughs, the ITCZ (Intertropical Convergence Zone), cold high-pressure systems extending to the south, and typhoons. The diveristy of the synoptic weather patterns among the chosen events help us examine the performance of our proposed "blended parameterization" method. Additionally, each event had rainfall exceeding 150 mm for 72 hours and affected large areas, making them representative of significant rainfall events that can have large impacts on life and production in Vietnam. These criteria ensure that the selected events are suitable for evaluating the effectiveness of the blended parameterization method under varying meteorological conditions}. In this subsection we will explain the rainfall totals and the synoptic weather patterns associated with these heavy rainfall occurrences summarized in {Table 1}.

\begin{table}[h!]
\caption{{Rainfall Events Summary}}
\centering
\resizebox{\textwidth}{!}{
\begin{tabular}{|l|p{1.8cm}|p{2.2cm}|p{2.5cm}|p{2cm}|p{5.5cm}|}
\hline
\textbf{Name} & \textbf{Date} & \textbf{Synoptic Weather Pattern} & \textbf{Rainfall Area} & \textbf{Rainfall Amount (mm)} & \textbf{Reason for Selection} \\ \hline
Event1         & May 21-24, 2022      & Low-pressure trough                   & Northern Vietnam             & 150-350                      & This event, driven by a low-pressure trough, tests the method's ability to handle severe rainfall due to low-pressure systems. \\ \hline
Event2         & July 9-12, 2022      & ITCZ and Southwest monsoon winds      & Central Vietnam              & 70-200                       & Demonstrates how the method handles rainfall caused by the interaction of ITCZ and monsoon winds. \\ \hline
Event3         & September 21-24, 2022 & Cold high-pressure system and ITCZ    & North and Central Vietnam    & 100-250                      & This event combines multiple weather systems, offering a challenging scenario for the model to predict. \\ \hline
Event4         & October 8-11, 2022    & Cold high-pressure system and Low-pressure trough & Central Vietnam & 70-500 & Tests how the model adapts to complex weather systems with both high and low-pressure areas affecting rainfall. \\ \hline
Event5         & October 14-17, 2022   & Typhoon Sonca and ITCZ                & Central Vietnam              & 150-700                      & This extreme event caused by a typhoon provides an opportunity to test the model's performance under severe, dynamic conditions. \\ \hline
\end{tabular}
}
\end{table}

Rainfall event from May 21 to May 24, 2022 (Event 1): The low-pressure trough responsible for the rainfall is positioned in the northern region of Central Vietnam, found between 18° and 20°N (Fig. S2a). Rainfall levels ranged from 150 mm to 350 mm in 72 hours (Fig. S2b).

Rainfall event from July 9 to July 12, 2022 (Event 2): Rainfall in the central region of Vietnam resulted from the southern edge of the ITCZ moving through the region, along with strong southwest monsoon winds (Fig. S3a). Accumulated 72-hour rainfall to 00z 12 July 2022 ranges from 70 mm to 150 mm, with some areas receiving over 200 mm (Fig. S3b)

Rainfall event from September 21 to September 24, 2022 (Event 3): The rainfall is caused by the southwestern boundary of the continental cold high-pressure system, combined with the ITCZ passing through the central part of Vietnam (Fig. S4a). The rainfall ranged from 100 mm to 200 mm, with some locations exceeding 250 mm (Fig. S4b).

Rainfall event from October 8 to October 11, 2022 (Event 4): The rain is caused by a continental high-pressure system extending into Vietnam and the northern edge of a low-pressure trough located at 7-10°N (Fig. S5a). Rainfall amounts commonly range from 70 mm to 150 mm in this region, with some areas receiving over 250 mm. In the southern part of Central Vietnam, rainfall is prevalent between 200 mm and 400 mm, with some areas receiving over 500 mm (Fig. S5b).

Rainfall event from October 14 to October 17, 2022 (Event 5): The rains were first caused by the circulation of Typhoon Sonca and then by the combination of the ITCZ with the low-pressure area that formed from the weakening of Typhoon Sonca. Rainfall generally ranges from 150 mm to 300 mm, with central Vietnam receiving 500 mm to 700 mm and some areas receiving over 700 mm (Figueres S6a and b).

\section{Method} \label{sec:Method}

\subsection{Operational multi-physics ensemble system in Vietnam}\label{sub:Operational multi-physics ensemble system in Vietnam}
To evaluate the effectiveness of our blended parameterization approach to forecast heavy rainfall events in Vietnam, it is essential to compare it with the operational forecasting system currently used by the Vietnam National Center for Hydro-Meteorological Forecasting (NCHMF). The Vietnam NCHMF utilizes a regional ensemble forecasting system based on the WRF model 
{\citep{Hung2023}}. The WRF model can provide up to 12,306,840 configurations from 23 microphysics schemes, 13 land surface schemes, 5 longwave radiation schemes, six shortwave radiation schemes, 14 planetary boundary layer schemes, seven surface layer schemes, and 14 cumulus schemes by selecting only one option in each physical process while the remaining options are not selected \citep{skamarock2019description}. 
{Based on this vast number of choices, the ensemble prediction system at Vietnam NCHMF consists of 32 (=$2^5$) forecast members, combined from two options for each of the five physical processes: cumulus convection, cloud microphysics, planetary boundary layer, shortwave radiation, and longwave radiation.}

\begin{itemize}
    \item The two options for the cumulus process are Kain-Fritsch and Tiedtke.
    \item The two options for the cloud microphysics process are WSM6 and Lin.
    \item The two options for the planetary boundary layer process are Shin-Hong and YSU.
    \item The two options for the shortwave radiation process are Dudhia and Goddard Shortwave.
    \item The two options for the longwave radiation process are RRTM and Goddard Longwave.
\end{itemize}

{The land surface and surface layer schemes, did not enable weighting parameters, using the Unified Noah Land Surface Model and Revised MM5 Scheme respectively}
 
\subsection{Multi-physics ensemble by blended parameterization} \label{sub:Multi-physics ensemble by blended parameterization}

Every physical parameterization computes the increment of state variables $\dfrac{dx}{dt}$, where $x$ is the state variable representing temperature, pressure, humidity, etc., and $t$ is time. Each option of the physical parameterization scheme $(Opt)$ represents the change in the state variables through a function $f(x, Opt)$, therefore:
\begin{equation}
    \frac{dx}{dt} = f(x, Opt_i)
    \label{eq:rate_of_change_x}
\end{equation}

Here, we represent the $i-th$ option among N options for the physical parameterization scheme. In this study, we focus on estimating the uncertainty in selecting physical parameterization schemes to improve the accuracy of the rainfall multi-physics ensemble forecast. Instead of choosing only one option for each physical parameterization scheme, we apply the blended parameterization approach \citep{mai2020simultaneously}. This approach parameterizes the selection of physical parameterization schemes using weighting parameters. Thus, it allows us to call all possible options of each physical parameterization simultaneously. The increment of state variables at each time step is calculated as the weighted average of the increments of all the selected parameterization scheme options. The blended parameterization approach reduces the model selection problem to parameter optimization. In the blended parameterization, Eq.(\ref{eq:rate_of_change_x}) is reformulated as:
\begin{equation}
    \frac{dx}{dt} = \sum_{i=1}^{N} w_i f(x, Opt_i)
    \label{eq:rate_of_change_sum}
\end{equation}

where $N$ is the total number of selected options of each physical parameterization scheme. These weights $w_i$ ($i$ = 1, …, $N$) must sum up to one:
\begin{equation}
    \sum_{i=1}^{N} w_i = 1
    \label{eq:weighting_parameters_sum}
\end{equation}

\begin{figure}[h]
 \centerline{\includegraphics[width=30pc]{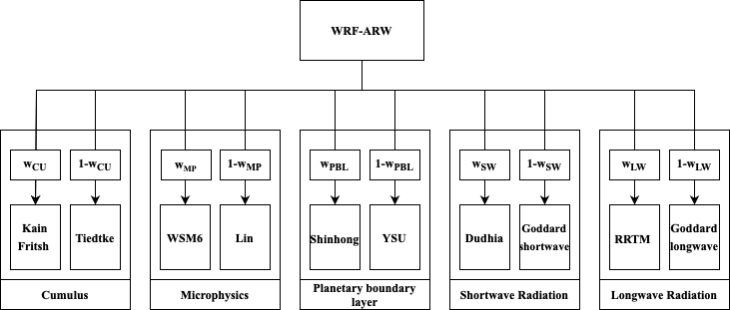}}
  \caption{Physical scheme selections and weighting parameters}\label{fig:Physical scheme selections and weighting parameters}
\end{figure}

Therefore, $N$-1 independent weighting parameters $w_i$ (i = 1, …, $N$-1) are introduced each between 0 and 1. In the study, the blended parameterization approach is applied to five physical parameterizations, and the selections for each physical parameterization scheme are the same as those of the Vietnam operational multi-physics ensemble system (Section \ref{sub:Operational multi-physics ensemble system in Vietnam}). Figure. \ref{fig:Physical scheme selections and weighting parameters} summarizes the configuration of our blended parameterization. The convection process will be weighted with  $w_{CU}$ for Kain Fritsh and 1-$w_{CU}$ for Tiedke. The microphysics process will be weighted with $w_{MP}$ for WSM6 and 1-$w_{MP}$ for Lin.  The planetary boundary layer process will be weighted with $w_{PBL}$ for Shinhong and 1-$(w_{PBL}$ for YSU. The shortwave process will be weighted with $(w_{SW}$ for Dudhia and 1-$w_{SW}$ for Goddard Shortwave. The longwave process will be weighted with $w_{LW}$ for RRTM and 1-$w_{LW}$ for Goddard Longwave. 

{From the view of the weighting parameters introduced in the blended parameterization, the Operational multi-physics ensemble system in Vietnam NCHMF assigns a weighting parameters ($w_{CU}$, $w_{MP}$, $w_{PBL}$, $w_{SW}$ and $w_{LW}$) to 0 or 1 to choose one of two schemes in each physical process. The state update is performed by a single selected and the other scheme is not called. The blended parameterization represents a more generalized framework. This new approach allows for the investigation of uncertainty in selecting physical parameterization schemes across a continuous spectrum of weighting values between 0 and 1. The proposed parameterization is largely different from the original operational system that assesses uncertainty over a discrete and countable model ensemble space. By introducing continuous weighting parameters, the blended parameterization address the limitations of the discrete model selection and enables to explore potentially more accurate and nuanced representations of physical processes through weighted averages of different physical parameterization schemes.}

\subsection{Optimizing weighting parameters in blended parameterization}\label{sub:Optimizing weighting parameters in blended parameterization}
{In this study the weighting parameters were optimized within the blended parameterization framework. Rather than identifying  a single optimal set of parameters, the goal is to determine the posterior probability distribution of these weights, which inherently accounts for the uncertainties in selecting physical parameterization schemes. This is achieved using the Markov Chain Monte Carlo (MCMC) method. A significant challenge in applying MCMC directly is the high computational cost, as evaluating the likelihood function requires running full WRF  for each proposed set of weighting parameters, which is expensive.  To address this, a surrogate model is used as an efficient estimator of this likelihood, avoiding the need for repeated run full WRF. The MCMC and surrogate model approach will be detailed in the following subsections (3.3.1 and 3.3.2).
}

\textit{3.3.1 {Markov Chain Monte Carlo}}

We estimate the posterior probability distribution of weighting parameters $\mathbf{w}$ = \{ $w_{CU}$, $w_{MP}$, $w_{PBL}$, $w_{SW}$, $w_{LW}$ \} by observation data. The posterior probability distribution can be computed by the Bayes' rule: 
\begin{equation}
    P(\mathbf{w}|\mathbf{O})\approx P(\mathbf{O}|\mathbf{w}) P(\mathbf{w})
    \label{eq:Bayesian}
\end{equation}
Where $P(\mathbf{w}|\mathbf{O})$ is the posterior probability distribution of $\mathbf{w}$ given by the observations $\mathbf{O}$, $P(\mathbf{O}|\mathbf{w})$ is the likelihood, and $P(\mathbf{w})$ is the prior probability distribution of $\mathbf{w}$. In this study, the prior $P(\mathbf{w})$ is assumed to be the bounded uniform distribution on [0,1]. The selection of the prior $P(\mathbf{w})$ signiﬁcantly affects the results of parameter optimization and uncertainty assessment (Sawada, 2020). Since there is no research on the application of weighting parameters in the selection of physical parameterization schemes, it is reasonable to assume that we do not have any information prior. The objective selection of the prior is outside the scope of this study. 
The posterior $P(\mathbf{w}|\mathbf{O})$ in Eq.(\ref{eq:Bayesian}) was obtained by the MCMC sampler, specifically through the Metropolis‐Hastings algorithm \citep{hastings1970monte}. First, the initial weighting parameter set $\mathbf{w}_0$ was specified. Then, the processes shown below were iterated: 

\begin{enumerate}
    \item For each iteration $i-th$, generate a new candidate weighting parameter set $\mathbf{w}_{Candidate}$ from the distribution $q(\mathbf{w}_{Candidate}|\mathbf{w}_i)$
    \item Compute the likelihood function $L(\mathbf{w}_{Candidate})$
    \item Calculate the acceptance ratio: $\alpha = \dfrac{L(\mathbf{w}_{\text{Candidate}})}{L(\mathbf{w}_i)}$
    \item Generate a random number b from the uniform distribution over [0, 1]. Then:
       \begin{itemize}
        \item If $b < \alpha$, accept the candidate parameter set and $\mathbf{w}_{i+1} = \mathbf{w}_{Candidate}$
        \item If $b \geq \alpha$, reject the candidate parameter set and $\mathbf{w}_{i+1} = \mathbf{w}_i$
    \end{itemize}
\end{enumerate}

In this study, $q(\mathbf{w}_{Candidate}|\mathbf{w}_i)$ was assigned to the Gaussian distribution, with the mean set to $\mathbf{w}_i$ and the standard deviation set to 0.01. The likelihood function $L$ is computed as follows:
\begin{equation}
    L = \prod_{k=1}^{M} \exp \left( -\frac{\left( \log(\mathbf{S}_k + \varepsilon) - \log(\mathbf{O}_k + \varepsilon) \right)^2}{2 \sigma_O^2} \right)
    \label{eq:Likelihood}
\end{equation}

This study divides the area into $M$ non-overlapping neighborhoods because rainfall simulations or forecasts often have high uncertainties about exact locations (Fig. S1). Considering the forecasts and observations in a neighborhood rather than at individual grid points reduces the impact of minor location errors. It allows for a more reasonable calculation of the likelihood function.  Each neighborhood is a square of $m$ x $m$ grid points; $m$ was set to 7 in this study. The value for each neighborhood is determined by the base ten logarithm of the sum of the rainfall values of all grid points within it. 
{$\mathbf{S}_k$ and $\mathbf{O}_k$ are the 72-hour accumulated simulated and observed rainfall, respectively, at each grid point in the $k$-th neighborhood.}
A tiny error $\varepsilon$ = 0.001 mm was added to avoid the rainfall in each neighborhood is equal to 0, making it impossible to calculate the base ten logarithms of the rainfall in the neighborhoods. $\sigma_O$ = 2 log(mm) is the observation error in the observed rainfall in this study. To simplify the acceptance rate calculation in practical terms, we will replace the likelihood function with a calculation using the log-likelihood function.
\begin{equation}
    \alpha = \exp \left( \left( \sum_{k=1}^{M} \frac{\left( \log(S_k + \varepsilon) - \log(O_k + \varepsilon) \right)^2}{-2 \sigma_O^2} \right)_{\text{Candidate}} - \left( \sum_{k=1}^{M} \frac{\left( \log(S_k + \varepsilon) - \log(O_k + \varepsilon) \right)^2}{-2 \sigma_O^2} \right)_{i} \right)
    \label{eq:acceptance}
\end{equation}

{The weighting parameters optimization in this study uses the MCMC method to obtain posterior probability distributions for weighting parameters rather than a single "best" parameter set. These distributions show the range of reasonable values and the associated uncertainty for each weighting parameter, with peaks indicating the most likely parameters. Then, the ensemble predictions are generated by sampling multiple sets of weighting parameters from these posterior distributions, which inherently accounts for the uncertainty in optimal parameterizations. Therefore, the study characterizes the probability distribution of plausible weighting parameters instead of defining a single fixed optimal parameter set.}

\textit{3.3.2 {Surrogate model}}

{In our study, the MCMC with Metropolis-Hastings algorithm is used to sample from the posterior probability distribution of the weighting parameters based on the likelihood function (Eq.5). However, evaluating the likelihood is computationally expensive because it requires running a full WRF model for each candidate a parameter set to simulate rainfall and compare it to observed values. Since the MCMC requires thousands or even millions of such evaluations to converge to the posterior distribution, directly using the expensive computational model for each step would be unrealistic \citep{sawada2020machine}. To address this, a surrogate model can be used as an efficient estimator. The surrogate model is trained to learn the relationship between the input weighting parameters and the output of log-likelihood. Before running MCMC, we generate a set of weighting parameter combinations and run the full WRF model to calculate the log-likelihood, creating the dataset of (parameter, log-likelihood) pairs to train the surrogate model. During MCMC iterations, instead of calculating log-likelihood by running the full WRF model, the surrogate model quickly calculates it by a proposed set of parameters with a cheaper computational cost. It allows the MCMC algorithm to explore the parameter space and sample from the posterior distribution within a reasonable computational time, overcoming the computational expense of running the full WRF model for each iteration. \citep{razavi2012review, sawada2020machine}. The key steps for constructing the surrogate model in our work are as follows:}
\begin{enumerate}
    \item Sample weighting parameters sets and run the WRF model for each set.
    \item Calculate the differences between the simulation and the observation using the log-likelihood by Eq.(\ref{eq:Likelihood})
    \item Use Gaussian process regression to learn the relationship between the weighting parameter sets and the log-likelihood to construct the surrogate model.
    \item Retrieve the posterior distribution {$P(\mathbf{w}|\mathbf{O})$}  by running the Metropolis-Hastings algorithm, in which the statistical surrogate model evaluates the log-likelihood.
\end{enumerate}

{Initially, we generated 1000 ensemble members of weighting parameters $\mathbf{w}$ utilizing quasi-Monte Carlo \citep{William1995}. Next, we computed the log-likelihood for these 1000 ensemble members by running the WRF model and comparing the simulated rainfall against observed data. This resulted in a dataset of 1000 pairs of $\mathbf{w}$ and log-likelihood. Following this, a surrogate model was developed using statistical machine learning that learned this dataset. Specifically, 900 of these pairs were used to train the surrogate model by the Gaussian process regression with the Rational Quadratic Kernel \citep{rasmussen2003gaussian}. Then, the remaining 100 sets of $\mathbf{w}$ and their corresponding 100 log-likelihood values were used to test the surrogate model’s ability to accurately mimic the relationship between the weighting parameters and the log-likelihood. With acceptable accuracy, the surrogate model can be used as a computationally efficient alternative to running the full WRF model within the MCMC algorithm.}

{Previous research has shown the effectiveness of multiobjective adaptive surrogate model-based optimization(MO-ASMO) techniques in optimizing parameters of numerical weather prediction (NWP) models to improve predictions \citep{chinta2020calibration,chinta2021assessment}. These adaptive methods iteratively refine the surrogate model based on optimization progress, potentially leading to higher accuracy. As we will discuss in Section 5, our study successfully developed the accurate surrogate model with no iterative processes, so that we did not use such an advanced method to construct surrogate models.}

\section{Experiment design} \label{sec:Experiment design}
\subsection{Experiment 1: Construction of the blended parameterization} 
\label{sub:Experiment 1: Construction of the blended parameterization}
This experiment aims to demonstrate the process of generating a new ensemble forecasting by the blended physical parameterization of the WRF model. {In Experiment 1, Event 3 was chosen randomly from the five heavy rainfall events section \ref{sub:study_cases}) to check if the ”blended parameterization” approach would be feasible. For this initial experiment, the specific characteristics of the chosen event are less critical than simply showing that the blending and optimization process can be successfully executed.} First, we sampled 1000 sets of 5 weighting parameters for cumulus convection, microphysics, planetary boundary layer, shortwave radiation, and longwave radiation processes using Quasi-Monte Carlo (QMC). Next, we ran the blended model using these 1000 sets of 5 weighting parameters to simulate one specific event.{ We then calculated the differences between simulated and observed rainfall amounts using log-likelihood, obtaining 1000 log-likelihood values between simulated and observed rainfall.} The surrogate model was constructed and tested to accurately mimic the relationship between the input weighting parameters and the output log-likelihood (section 3.3.2). 
Then, the surrogate model can be used as a computationally efficient alternative to running the full WRF model within the MCMC algorithm to obtain posterior distribution. By sampling from this posterior distribution, we generated a 32-member ensemble forecasting system conditioned by rainfall observation in the target event, which is also Event 3. Finally, we confirmed that our blended parameterization effectively reduces the error in rainfall using the same rainfall data of that event as those used for training weighting parameters. 
 
\subsection{Experiment 2: Forecasting capability of the blended parameterization} 
\label{Experiment 2: Forecasting capability of the blended parameterization}
\begin{figure}[h]
 \centerline{\includegraphics[width=25pc]{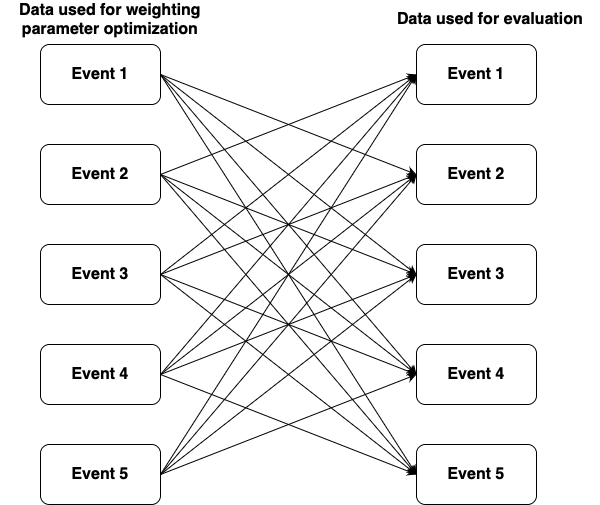}}
  \caption{The data used to optimize the weighting parameters and evaluate the performance of the new system in Experiment 2}\label{fig:Experiment 2}
\end{figure}

The WRF model with the blended parameterization is expected to be applied in operational forecasting in Vietnam. In this sense, it is necessary to evaluate the capability of the blended parameterization to forecast future heavy rainfall events which are not included in the training data. The objective of Experiment 2 is to investigate the rainfall forecasting capability of the new ensemble forecasting system based on blended parameterization. We assessed the skill to simulate rainfall events in which rainfall observation were not used to optimize the weighting parameter. In Experiment 2, we used a cross-validation method, where each of the five heavy rainfall events (Section \ref{sub:study_cases}) was sequentially used for weighting parameter optimization, while the remaining four events were used for performance evaluation (Fig. \ref{fig:Experiment 2}). This approach comprehensively assessed the system's generalization ability, minimizing the risk of overfitting to any single event. 
\subsection{Experiment 3: The blended parameterization optimized by multiple heavy rainfall events}
\label{Experiment 3: The blended parameterization optimized by multiple heavy rainfall events}
\begin{figure}[h]
 \centerline{\includegraphics[width=25pc]{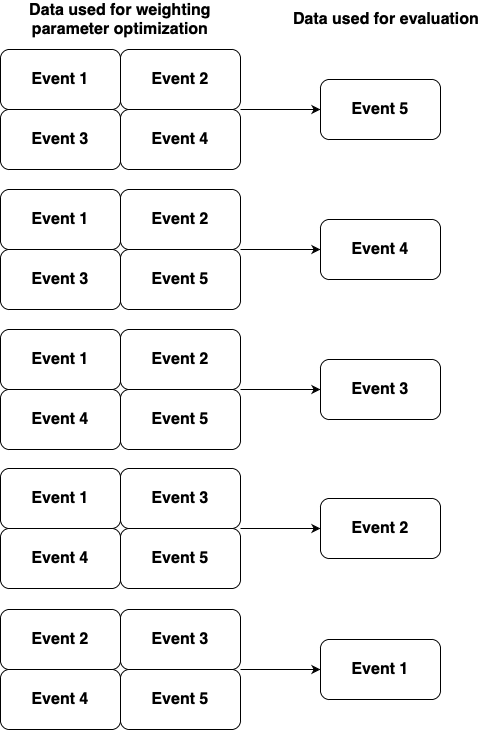}}
  \caption{Same as Fig. 3 but for Experiment 3}\label{fig:Experiment 3}
\end{figure}

While previous experiments relied on a single rainfall event to optimize weighting parameters, Experiment 3 was designed to evaluate the effectiveness of incorporating broader datasets from multiple rainfall events in optimizing the weighting parameters. We used rainfall data from four distinct events to estimate the weighting parameters and subsequently simulate rainfall for a fifth event (Fig. \ref{fig:Experiment 3}). The performance of the blended parameterization system in simulating heavy rainfall in Experiment 3 was compared with that of the operational system.

\section{Result} \label{sec:Result}
\subsection{Experiment 1: Construction of the blended parmeterization} \label{sub:result Experiment 1: Construction of the blended parameterization}
\begin{figure}[h]
 \centerline{\includegraphics[width=25pc]{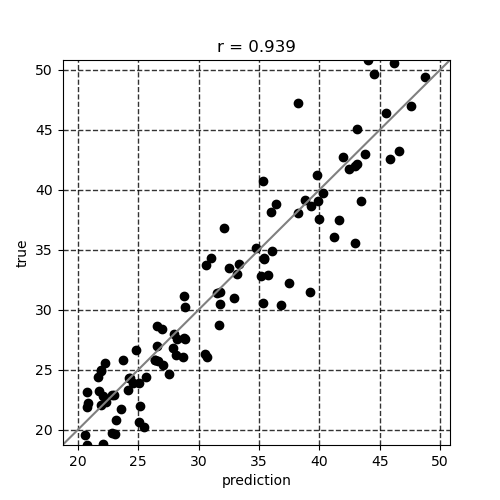}}
  \caption{Comparison between the true log-likelihood calculated by the Blended model and the observed rainfall (vertical axis), and the predicted log-likelihood calculated by the surrogate model, which is developed using the rainfall data from Event 3 (horizontal axis)}\label{fig:Surrogate_model_exp1}
\end{figure}

We drove the WRF model with the blended parameterization with the 1000 ensembles of the weighting parameters randomly drawn by the QMC sampling. Then, the statistical machine learning‐based surrogate model was constructed using Gaussian process regression, {trained on dataset 900 weighting parameters and their corresponding log-likelihood values.} {To evaluate the accuracy of the constructed surrogate model, we utilized the 100 remaining sets of weighting parameters and log-likelihood values that were not used in the training process (Section \ref{sub:Optimizing weighting parameters in blended parameterization}.2). These 100 sets of 5 weighting parameter values were used as input into the trained surrogate model to obtain predicted log-likelihood values. 
Figure \ref{fig:Surrogate_model_exp1} compares the predicted log-likelihood values (horizontal axis) with the "true" values (vertical axis) calculated using the WRF model with the blended parameterization for Event 3, with each point representing a test case from one of the 100 untrained weighting parameter sets
The correlation coefficient (r) is 0.939 indicating the accuracy of surrogate model. Therefore, it can be justified to use this surrogate model for parameter optimization and uncertainty assessment.} 

\begin{figure}[H]
 \centerline{\includegraphics[width=40pc]{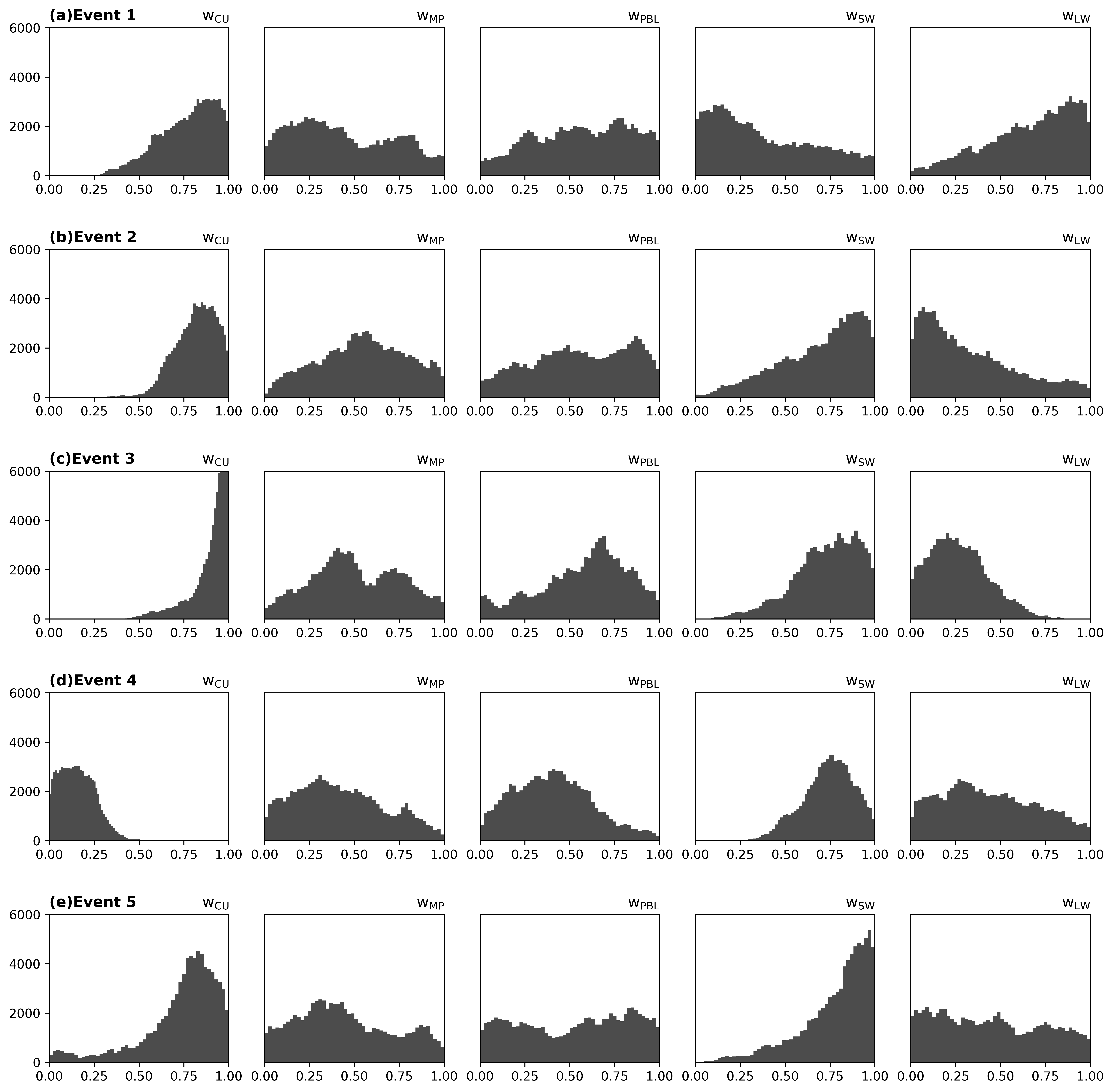}}
  \caption{MCMC-sample posterior probability distributions of the weighting parameters based on the observed rainfall of five events. From top to bottom, the events are: (a) Event 1; (b) Event 2; (c) Event 3; (d) Event 4 and (e) Event 5. Each event consists of five distributions. From left to right the plots represent the distributions for $w_{CU}$, $w_{MP}$, $w_{PBL}$, $w_{SW}$ and $w_{LW}$. In each distribution plot, the horizontal axis shows weighting parameters, and the vertical axis shows their frequencies.}\label{fig:mcmc_Exp2}
\end{figure}

Figure \ref{fig:mcmc_Exp2}c shows the posterior distributions sampled by the $10^5$ iterations of the MCMC The log-likelihood is calculated by the surrogate model. The plots from left to right are the distributions of $w_{CU}$, $w_{MP}$, $w_{PBL}$, $w_{SW}$ and $w_{LW}$ respectively. The posterior distributions do not follow the Gaussian distribution, which shows that the nonparametric approach by MCMC is appropriate. The posterior probability distributions for five weighting parameters during Event 3 show distinct patterns. For $w_{CU}$, there is a pronounced skew towards 1.0, indicating a greater contribution from the Kain-Fritsch scheme, while the contribution from the Tiedke scheme (values near to 0) is small. The distribution of $w_{MP}$ shows two distinct peaks around 0.3 and 0.7 and many combinations of WSM6 ($w_{MP}$=1.0) and Lin ($w_{MP}$=0.0) are accepted. The distribution for $w_{PBL}$ shows that many combinations between 0.4 and 0.8 are sampled, with a peak around 0.7, indicating a slightly greater contribution from the Shin-Hong scheme ($w_{PBL}$=1.0) compared to YSU ($w_{PBL}$=0.0). For $w_{SW}$, the distribution is strongly skewed towards 1.0 with a peak around 0.8, indicating a greater contribution from the Dudhia scheme ($w_{SW}$=1.0) against Goddard. Finally, the distribution of $w_{LW}$ is skewed towards 0.0, with a peak around 0.2, indicating a greater contribution from the Goddard Longwave scheme ($w_{LW}$=0.0) compared to RRTM ($w_{LW}$=1.0). 

We drove WRF with this optimized blended parameterization. We use the 32 ensemble members of the weighting parameters drawn from the posterior distributions shown in Fig. \ref{fig:mcmc_Exp2}c. Then, we compared it with the simulation driven by the Operational system which is realized by setting the weighting parameters 0 or 1. {The RMSE in logarithmic scale ($RMSE_{log}$) is used to assess the performance of our optimized blended parameterization in comparison to the Operational System:}

{
\begin{equation}
     RMSE_{log} = \sqrt{\frac{1}{M} \sum_{k=1}^{M} 
     \frac{\left( \log(S_k + \varepsilon) - \log(O_k + \varepsilon) \right)^2}
     {-2 \sigma_O^2}}
     \label{eq:rmselog}
\end{equation}
}

\begin{figure}[H]
 \centerline{\includegraphics[width=25pc]{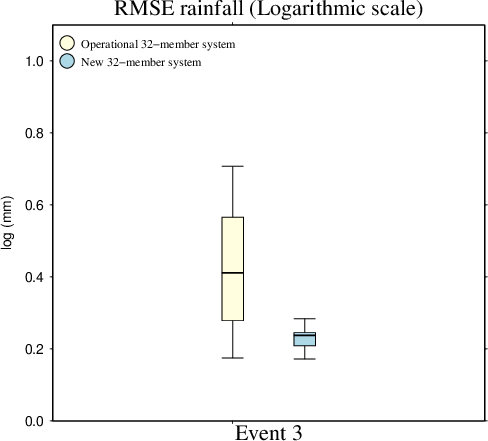}}
  \caption{Boxplots of RMSE in logarithmic scale between observation and simulation rainfall by the operational 32-member system (yellow) and the new 32-member system (blue) that drawn from the posterior distribution for the rainfall for Event 3.}\label{fig:rmse_Exp2}
\end{figure}

{
In Eq.7, $S_k$ and $O_k$ refer to the 72-hour accumulated rainfall that sums at grid points in a $k$-th non-overlapping neighborhood (Section \ref{sub:Optimizing weighting parameters in blended parameterization}.1). This calculation is done for each member of the ensemble prediction system. Figure \ref{fig:rmse_Exp2} shows the distribution of calculated $RMSE_{log}$ across the 32 members of each system: the operational (yellow) and new blended parameterization (blue) for Event 3. This figure demonstrates that the blended parameterization substantially outperforms the operational ensemble prediction system, and the construction of the blended model system is successful and reasonable.}

{The blended parameterization approach incorporates contributions from all selected physical parameterization schemes, with their influence represented by their weighting parameters. This creates a new system capable of improving rainfall forecasts compared to the operational system. In the blended parameterization, the higher the weighting parameter value assigned to a scheme, the more its representation of the physical processes influences the overall simulation. When this influence resulted in a better agreement with observations (quantified by the likelihood function), that scheme received a higher posterior probability and thus a higher value weighting parameter in the ensemble. These schemes simulate more accurately the key physical processes governing convection development and precipitation under the specific atmospheric conditions of the simulated events.
For instance, if the Kain-Fritsch cumulus scheme received a higher weighting parameter value for a heavy rainfall event (Figure 6c), it suggests that its formulation of convection (e.g. clouds forming, developing, and precipitating) was more realistic compared to the Tiedke scheme for that particular heavy rainfall event. A higher weighting parameter value for a specific radiation scheme (like Dudhia in Figure 9) indicates its ability to model the radiative fluxes and their interaction with clouds and the surface during heavy rainfall events that lead to a more accurate atmospheric heating and cooling rates, affecting precipitation. Planetary boundary layer (PBL) schemes model the exchange of momentum, heat, and moisture between the surface and the free atmosphere. A particular PBL scheme with higher weight might simulate surface fluxes and turbulent mixing accurately that influence the rainfall by controlling the supply of moisture and instability from the lower atmosphere. In summary, the physical parameterization scheme with higher weighting parameter value indicates that its formulation and the physical processes it represents were more aligned with the real atmosphere during the simulated heavy rainfall events leads to more accurate precipitation simulations.} In the next section, we will validate the skill of this ensemble rainfall prediction system using rainfall data which were not used for the optimization.

\subsection{Experiment 2: Forecasting capability of the blended parameterization} \label{sub: result Experiment 2: Forecasting capability of the}
As we demonstrated in Experiment 1, we successfully mimicked the relationship between the weighting parameters and log-likelihood in the other events. Our surrogate models can reproduce log-likelihood brought by new weighting parameters with correlation coefficients higher than 0.7  (Fig. S7). The posterior distributions of the five weighting parameters based on Events 1, 2, 4, and 5 are shown in Fig. \ref{fig:mcmc_Exp2}a, b, d, and e, respectively. It can be observed that the posterior probability distributions of $w_{CU}$ show Kain-Frisch is preferred in Events 1,2 and 5 while Tiedke is preferred in Event 4. It implies that the nature of the bias in the WRF model for Event 4 differs from that in the other events. 
The $w_{MP}$ distributions shown in the second plots of Fig. 6a, b, d and e indicate the spanning, which suggests that many combinations between Lin and WSM6 schemes are accepted. Lin is slightly preferred in Events 1, 4, and 5, but there is a balance between Lin and WSM6 in Event 2. Although the model tends to favor Lin overall, both schemes can be applied consistently across all events. 
The $w_{PBL}$ distribution shows Shin-Hong is preferred in Events 1, 2 and 5, while YSU is preferred in Event 4. This observation is consistent with  $w_{CU}$ distributions, where Event 4 displays a particularly distinctive posterior compared to the other rainfall events.  This again suggests that the bias characteristics in the WRF model during this event differ from those observed in the remaining events. 
The fourth plot in Fig. 6a shows the distribution of $w_{SW}$  skew to 0.0, which means Goddard is preferred in Event 1, while the fourth plots of Fig. 6b, d, and e illustrate that Dudhia is preferred in three Events 2, 4, and 5. 
The distribution of $w_{LW}$ tends to be similar to $w_{SW}$  when RRTM dominates in Event 1 while Goddard Longwave is more favourited in the remaining three events. However, it is not so evident in the two Events 4 and 5 as seen in the distribution of $w_{SW}$. This suggests that the nature of the shortwave and longwave radiation processes in Event 1 differs from those in the other events. 

\begin{figure}[H]
    \begin{minipage}{0.35\textwidth}
        \centering
        \includegraphics[width=\linewidth]{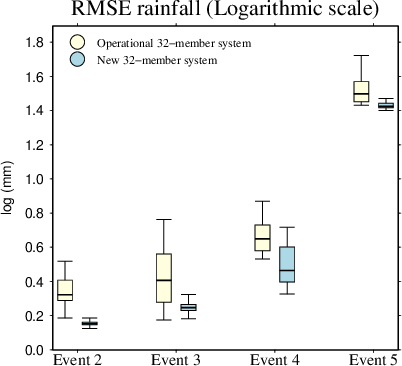}
        \subcaption{} \label{fig:exp2_rmse_2022052100_logscale_20240723.png}
    \end{minipage}
    \hfill
    \begin{minipage}{0.35\textwidth}
        \centering
        \includegraphics[width=\linewidth]{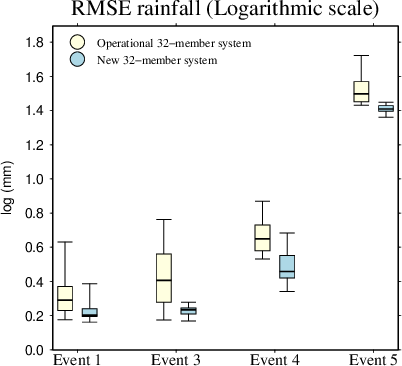}
        \subcaption{} \label{fig:exp2_rmse_2022070900_logscale_20240723.png}
    \end{minipage}

    \begin{minipage}{0.35\textwidth}
        \centering
        \includegraphics[width=\linewidth]{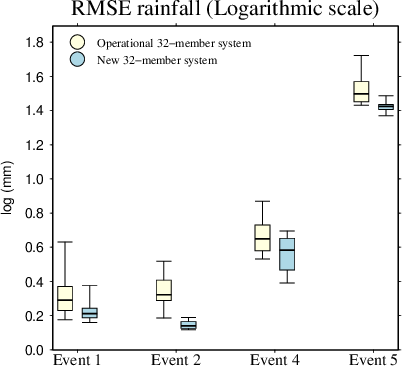}
        \subcaption{} \label{fig:exp2_rmse_2022092100_logscale_20240723.png}
    \end{minipage}
    \hfill
    \begin{minipage}{0.35\textwidth}
        \centering
        \includegraphics[width=\linewidth]{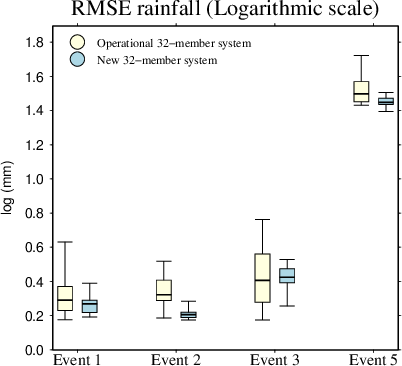}
        \subcaption{} \label{fig:exp2_rmse_2022100800_logscale_20240723.png}
    \end{minipage}

    \begin{minipage}{0.35\textwidth}
        \centering
        \includegraphics[width=\linewidth]{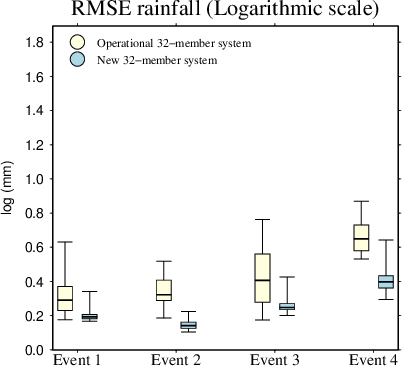}
        \subcaption{} \label{fig:exp2_rmse_2022101400_logscale_20240723.png}
    \end{minipage}

    \caption{Boxplots of RMSE in logarithmic scale between observation and simulation by the operational 32-member system (yellow) and the new 32-member system (blue) that were drawn from the posterior distribution based on rainfall data from (a) Event 1, (b) Event 2, (c) Event 3, (d) Event 4, and (e) Event 5 for the four remain rainfall events.}
    \label{fig:all_rmse_exp2_images}
\end{figure}

Five new 32-member forecasting systems were generated by sampling the set of five weighting parameters obtained from the posterior distribution. Unlike Experiment 1 (section 5.1), these new ensemble forecasting systems are evaluated using the remaining events. The simulation results are compared with the operational ensemble system. {Figure \ref{fig:all_rmse_exp2_images} presents boxplots comparing the $RMSE_{log}$ between observed and simulated rainfall for five heavy rainfall events. 
This is a cross-validation approach to assess the forecasting capability on unseen events.} {The key observation is that the new blended parameterized system (blue) generally shows improved forecasting capability with lower $RMSE_{log}$ values compared to the operational system (yellow) for most rainfall events.  The blue boxplots tend to have a lower overall range (smaller total spread) and median of $RMSE_{log}$ than the corresponding yellow boxplots (Fig. 8a, b, c, and e). This indicates that optimizing the weighting parameters, even with data from a single event, can lead to more skillful rainfall forecasts for other events.} {However, the ensembles optimized by Event 4 (Fig. 8d) did not significantly improve the forecasts for Events 1 and 3. This suggests that the optimal blending of physical schemes might be event-dependent, possibly due to different meteorological characteristics or model biases in Event 4, as indicated by its distinct cumulus convection  posterior distributions in Figure 6d.} {Overall, Figure 8 demonstrates the potential of the blended parameterization to enhance rainfall forecasting for severe storms beyond the event used for optimization.}

\subsection{Experiment 3: The blended parameterization optimized by multiple heavy rainfall events} \label{sub: result Experiment 3: The blended parameterization optimized by multiple heavy rainfall events}
\begin{figure}[H]
 \centerline{\includegraphics[width=40pc]{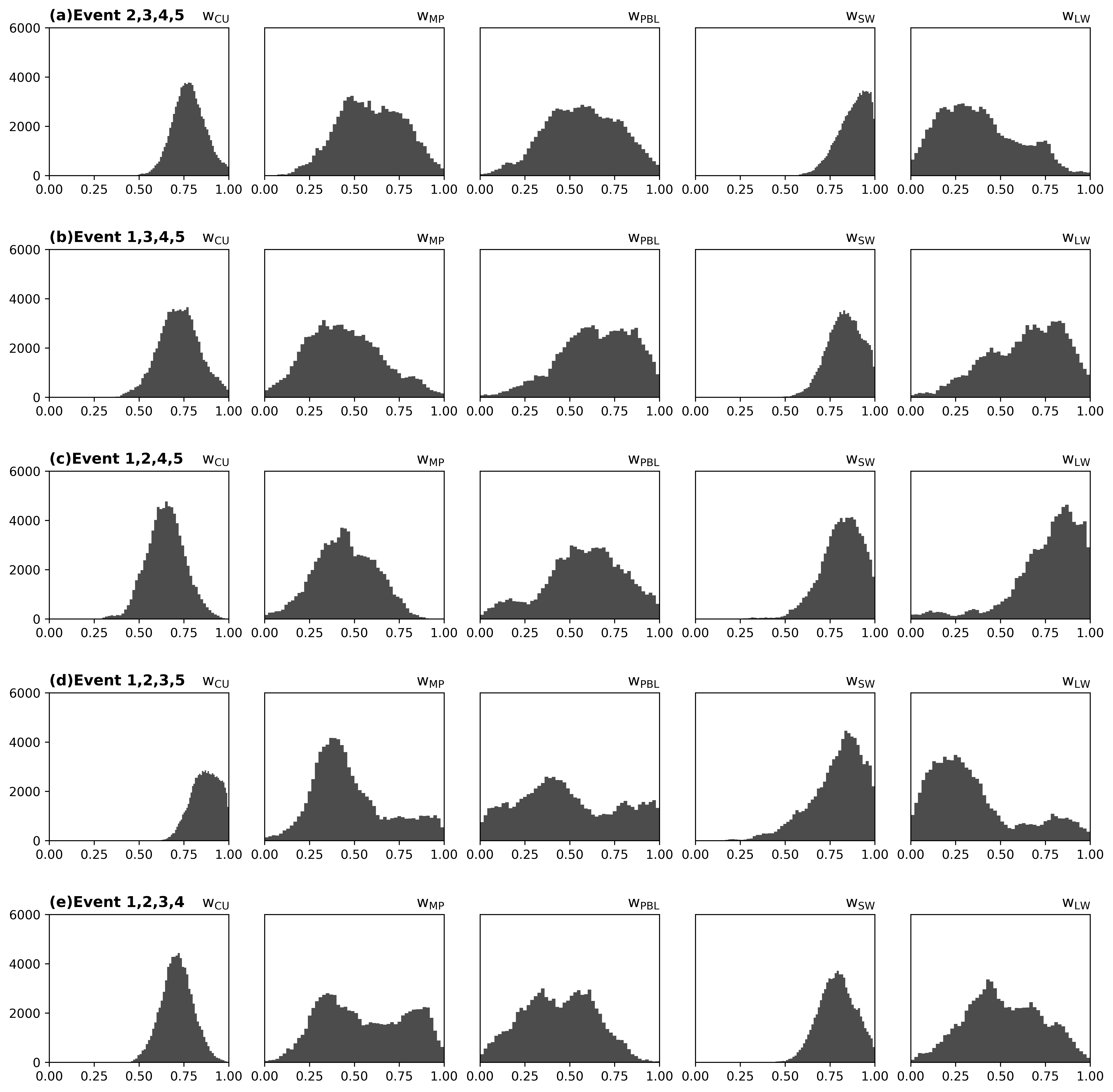}}
  \caption{Same as Fig. 6 but from top to bottom, the events are: (a) Event 2,3,4,5; (b) Event 1,3,4,5 (c); Event 1,2,4,5; (d) Event 1,2,3,5 and (e) Event 1,2,3,4}\label{fig:mcmc_Exp3}
\end{figure}

Even when multiple rainfall events are used to calculate log-likelihood, our surrogate model can accurately mimic the response of WRF with correlation coefficients higher than 0.7 (Fig. S8), except for the surrogate in Fig. S8c, which shows slightly lower r = 0.693. 
{The posterior distributions of the weighting parameters were obtained from different combinations of rainfall events (Fig. \ref{fig:mcmc_Exp3}a–e). The weighting parameters optimized separately for each combination. Each panel represents a posterior distribution derived from a unique combination of four out of the five available heavy rainfall events.}
Across all rainfall events, the Cumulus scheme is predominantly led by the Kain-Fritsch scheme, indicating its significant role in the model’s convective processes. 
Similarly, Dudhia always plays a dominant role in the Shortwave scheme, with most of the values of $w_{SW}$ concentrated between 0.4 and 1.0, indicating its significant impact on the shortwave radiation in all cases. Meanwhile, the Longwave process shows different influences between the Goddard Longwave and RRTM. Goddard Longwave dominates when considering the combined rainfall of Events 2, 3, 4, 5 (Fig. \ref{fig:mcmc_Exp3}a) and Events 1, 2, 3, 5 (Fig. \ref{fig:mcmc_Exp3}d) while RRTM plays a more significant role when considering the combined observations from Events 1, 3, 4, 5 and 1, 2, 4, 5 (Fig. \ref{fig:mcmc_Exp3}b and c). In the remaining case, the contributions from both schemes are more balanced, indicating an influence change compared to the other periods. The PBL scheme is mainly driven by the Shin-Hong scheme using rainfall in 3 cases (Fig. \ref{fig:mcmc_Exp3}a, b and c) while YSU takes over as the major contributor in two cases (Fig. \ref{fig:mcmc_Exp3}d and e). Regarding the Microphysics scheme, there is a more balanced contribution between the WSM6 and Lin schemes. However, there is a slight preference for WSM6 in the case of combined rainfall data of Events 2, 3, 4, 5 while if using rainfall from Events 1, 3, 4, 5 and 1, 2, 4, 5 there is a greater bias towards Lin. In summary, while the Kain-Fritsch and Dudhia schemes are the leading options in their respective schemes, the preferred choices of the Longwave, microphysical, PBL processes vary depending on the specific rainfall event.

\begin{figure}[H]
    \begin{minipage}{0.4\textwidth}
        \centering
        \includegraphics[width=\linewidth]{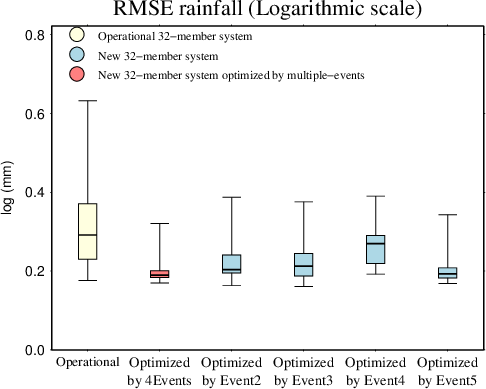}
        \subcaption{} \label{fig:exp3_rmse_2022052100_logscale_20240731_fix.png}
    \end{minipage}
    \hfill
    \begin{minipage}{0.4\textwidth}
        \centering
        \includegraphics[width=\linewidth]{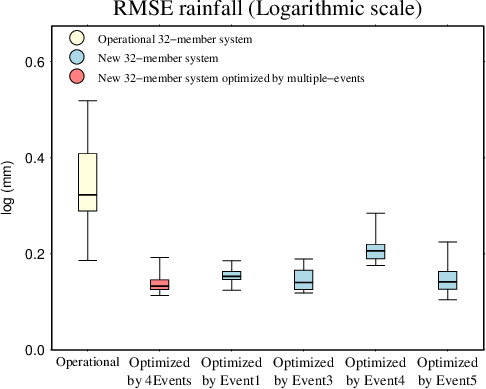}
        \subcaption{} \label{fig:exp3_rmse_2022070900_logscale_20240731_fix.png}
    \end{minipage}

    \begin{minipage}{0.4\textwidth}
        \centering
        \includegraphics[width=\linewidth]{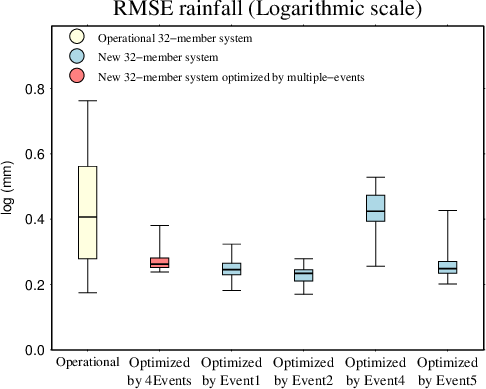}
        \subcaption{} \label{fig:exp3_rmse_2022092100_logscale_20240731_fix.png}
    \end{minipage}
    \hfill
    \begin{minipage}{0.4\textwidth}
        \centering
        \includegraphics[width=\linewidth]{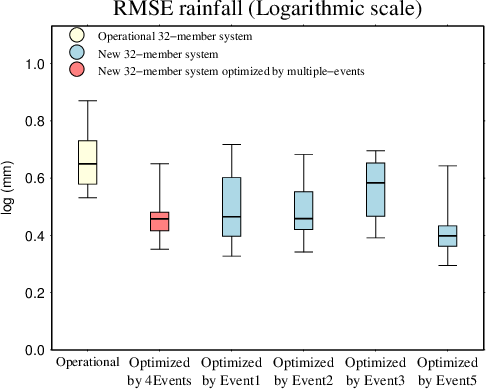}
        \subcaption{} \label{fig:exp3_rmse_2022100800_logscale_20240731_fix.png}
    \end{minipage}

    \begin{minipage}{0.4\textwidth}
        \centering
        \includegraphics[width=\linewidth]{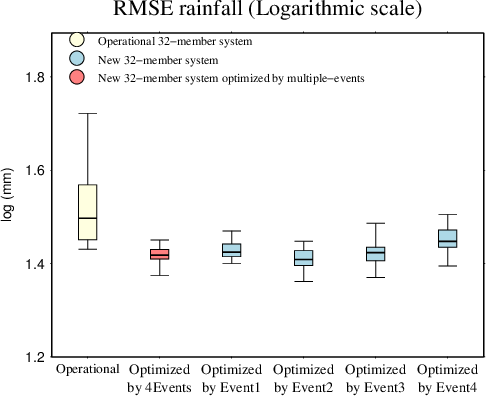}
        \subcaption{} \label{fig:exp3_rmse_2022101400_logscale_20240731_fix.png}
    \end{minipage}    

   \caption{Boxplots of RMSE in logarithmic scale among the operational system (yellow), the new ensemble system optimized by multiple-events (red), and the new ensemble systems optimized by single-events (blue) and then forecasted for the rainfall events on (a) Event 1; (b) Event 2; (c) Event 3; (d) Event 4; (e) Event5} \label{fig:all_exp3_rmse_images}
\end{figure}

{Figure \ref{fig:all_exp3_rmse_images} extends the evaluation by comparing the performance when the blended parameterization is optimized using rainfall data from multiple events. Similar to Figures 6 and 8, it displays boxplots of $RMSE_{log}$ for the five rainfall events. The yellow boxplots again represent the operational system. 
The red boxplots represent the performance of ensemble systems where the weighting parameters were optimized using the combined rainfall data of four of the five events, and then used to simulate the remaining single event. The blue boxplots show the performance of the single-event optimized systems from Experiment 2 (Fig. 8).
The most significant finding is that the blended parameterized system optimized by multiple events (red) often exhibits lower $RMSE_{log}$ values than the systems optimized by a single event (blue). This is evident in several of the subfigures (e.g., Figs. \ref{fig:all_exp3_rmse_images}a, b, c, e), suggesting that using a broader range of rainfall event data leads to a more stable and generally better estimation of the weighting parameters. This multi-event optimization appears to improve the robustness of the blended parameterization across different severe weather scenarios.
Furthermore, the variance of the forecasted rainfall, as indicated by the spread of the boxplots, is also generally reduced for the multi-event optimized systems (red) compared to the single-event optimized systems (blue). Consistently, the operational system (yellow) shows the highest $RMSE_{log}$ across all five events, confirming that the blended parameterization approach is more effective in reducing rainfall forecast errors. Figure 10 highlights the benefit of incorporating information from multiple heavy rainfall events to achieve a more reliable and accurate blended parameterization for operational forecasting.}


\section{{Conclusion and Discussion}}
This study introduced a new method to overcome the limitations of the NWP system in assessing and mitigating the uncertainty in selecting physical parameterization schemes \citep{wang2021quantifying, mai2020simultaneously}. We reformulated the WRF model to parameterize the selection of the physical parameterization schemes for cloud microphysics, convection, planetary boundary layer physics, shortwave radiation, and longwave radiation with their corresponding weighting parameters. New physics options for the model are defined by blending the physical scheme options from the discrete model. This converts the finite and discrete model ensemble space into an infinite and continuous model space \citep{mai2020simultaneously} and reduces the discrete model selection problem to a parameter optimization problem. {While Mai et al.(2020) pioneered the concept of blended or weighted selecting process options for sensitivity analysis in hydrological modeling, our study marks the first application of a similar blended parameterization idea to an atmospheric model with the specific objective of enhancing severe rainfall prediction.}

{The new aspect of our blended parameterization approach has demonstrated improvements when compared to other methods of quantifying uncertainty, such as Bayesian Model Averaging (BMA). In our blended parameterization, we used MCMC to estimate the posterior probability distribution of weighting parameters that determine the contribution of the different physical parameterization schemes within the WRF model, conditioned by rainfall observations. Similarly, BMA relies on Bayes' rule to determine the posterior probabilities, which act as weighting parameters, for different candidate models \citep{Sloughter2007, Zhu2015}. 
However, crucial key differences exist between these two approaches. Our blended parameterization weight the selection of different physical parameterization schemes within a single model (WRF). The fundamental structure of the WRF model remains the same, but its internal physics are determined by a blend of schemes, effectively transforming the discrete model ensemble space into an infinite and continuous model space.  In contrast, BMA weights the predictions from entirely different and discrete models \citep{Sloughter2007}, which could be different versions of WRF with distinct physics configurations or even entirely different forecast models altogether, making the model space in BMA typically discrete and finite \citep{Javanshiri2021}. 
The primary goal of our blended parameterization is to quantify the uncertainty in selecting physical parameterization scheme by allowing the model to simultaneously utilize a combination of them, with the optimal blending determined based on posterior distribution of weighting parameter by MCMC. BMA's main goal is to account for model structural uncertainty by combining predictions from different plausible models. Given that our research focused on exploring the continuous space of physics parameterization combinations within the WRF model to improve severe storm ensemble prediction in Vietnam, BMA, which considers a discrete set of distinct models, was not the direct approach suited for this objective. Our blended parameterization effectively reformulated the model selection problem into a continuous parameter optimization problem solved by MCMC, allowing for a clearer exploration of the uncertainties in physical parameterization within a single model framework.}

We successfully optimized the weighting parameters of the blended parameterization with the reasonably small computational cost using a surrogate model. The newly developed blended parameterization-based WRF model has been applied to five severe rainfall events and significantly outperforms the Vietnam NCHMF’s operational ensemble prediction system. {Note that the analysis of the posterior distributions of the weighting parameters revealed the relatively more influential physical parameterization schemes for the studied events. For instance, in Experiment 1 (Event 3), the Kain-Fritsch scheme for cumulus convection, the Dudhia scheme for shortwave radiation, and the Goddard Longwave scheme for longwave radiation showed greater contributions based on the higher weighting parameter values. Similarly, Experiment 3, which optimized weighting parameters using multiple rainfall events, indicated a consistent preference for the Kain-Fritsch scheme for cumulus and the Dudhia scheme for shortwave radiation across most cases. However, it is crucial to understand that the specific schemes with larger weighting parameters imply that their physical characteristics align better with the observations than the other options. The blended parameterization approach incorporates contributions from all considered physical parameterization schemes, with their influence being proportional to the optimized values of their respective weighting parameters. This blending allows for a more nuanced representation of physical uncertainties than a discrete selection of a single scheme for each physical process.}

There are several limitations in the present work. First, the blended parameterized atmospheric model increases the computational cost since it calls more physical parameterization schemes at the same time, although they can be efficiently parallelized. 
{Once the optimization of the weighting parameters is completed, the computational cost of the new ensemble prediction system will increase,  but it will not exceed twice that of the original physics ensemble of the operational system in Vietnam. Although two physical parameterization schemes are called, this does not mean the computational cost is exactly twice as much. This is because the blended parameterization system involves costs associated with dynamic processes, parameterizations that are not blended, and other overhead processes. Moreover, since the two parameterization calls can be parallelized, the computational time is not necessarily doubled. We consider that this increase in computational cost is operationally feasible in many cases.}

Second, this work considers only the uncertainties in the selection of physical parameterizations although there are many other sources of uncertainties in NWP systems as raised in Section\ref{sec:Introduction}. Our ensemble rainfall forecasting might be overconfident since we do not explicitly consider uncertainties in input data and model parameters. It is necessary to consider uncertainties in initial and boundary conditions, and model parameters in addition to model structural uncertainties that we considered in this work. Future work should focus on considering all the relevant uncertainties in rainfall probabilistic forecasting by combining our proposed blended parameterization method and the existing ensemble-based data assimilation \citep{sawada2024efficient}.

Third, it is more difficult to interpret the simulation results from the blended parameterized model than those from conventional atmospheric models. Although the weighted average of the increments from multiple parameterizations conserves mass and energy and each one of schemes are physically consistent, the blended parameterization is not necessarily physically consistent. Future work should focus on investigating mechanism of the improvement of rainfall forecast by the blended parameterizations.

\clearpage
\acknowledgments

This work was supported by the JSPS RONPAKU program (Grant R12201) and the JST Moonshot R \& D program (Grant JMPJMS2281).
Khanh Hung Mai was partly supported by the Ministry of Science and Technology's project titled: "Research and application of artificial intelligence technique to build the forecasting system of tropical cyclone activity over Bien Dong Sea and affecting Vietnam with lead time up to 3 days” (code: KC-4.0-46/19-25). This work was also supported by the Grants-in-Aid for Scientific Research (KAKENHI) project (Grant 24KK0073: Research on Radar Data Assimilation in Tropical and Subtropical Regions for Improving Precipitation Forecast in Vietnam).

%
%
\datastatement
All Vietnam observation data related to this paper can be requested from the authors for research purposes. 
The Final Global Analysis (FNL) data are stored in the NCAR RDA (Research Data Archive) at the National Center for Atmospheric Research, Computational and Information Systems Laboratory
https://rda.ucar.edu/datasets/d083003/dataaccess/


%






%




\begin{thebibliography}{41}
\providecommand{\natexlab}[1]{#1}
\providecommand{\url}[1]{\texttt{#1}}
\renewcommand{\UrlFont}{\rmfamily}
\providecommand{\urlprefix}{URL }
\expandafter\ifx\csname urlstyle\endcsname\relax
  \providecommand{\doi}[1]{https://doi.org/\discretionary{}{}{}#1}\else
  \providecommand{\doi}{https://doi.org/\discretionary{}{}{}\begingroup \urlstyle{rm}\Url}\fi
\providecommand{\eprint}[2][]{\url{#2}}

\bibitem[{Afshar et~al.(2020)Afshar, Azadi,, and Rezazadeh}]{afshar2020uncertainty}
Afshar, M.~A., M.~Azadi, and M.~Rezazadeh, 2020: Uncertainty reduction in quantitative precipitation prediction by tuning of kain--fritch scheme input parameters in the wrf model using the simulated annealing optimization method. \textit{Meteorological Applications}, \textbf{27~(4)}, e1919.

\bibitem[{Chen et~al.(2012)Chen, Yen, Tsay, Thanh,, and Alpert}]{Chen2012}
Chen, T.~C., M.~C. Yen, J.~D. Tsay, N.~T.~T. Thanh, and J.~Alpert, 2012: Synoptic development of the hanoi heavy rainfall event of 30-31 october 2008: Multiple-scale processes. \textit{Monthly Weather Review}, \textbf{140}, 1219--1240, \doi{10.1175/MWR-D-11-00111.1}.

\bibitem[{Chinta and Balaji(2020)Chinta, and Balaji}]{chinta2020calibration}
Chinta, S., and C.~Balaji, 2020: Calibration of wrf model parameters using multiobjective adaptive surrogate model-based optimization to improve the prediction of the indian summer monsoon. \textit{Climate Dynamics}, \textbf{55~(3)}, 631--650.

\bibitem[{Chinta et~al.(2021)Chinta, Yaswanth~Sai,, and Balaji}]{chinta2021assessment}
Chinta, S., J.~Yaswanth~Sai, and C.~Balaji, 2021: Assessment of wrf model parameter sensitivity for high-intensity precipitation events during the indian summer monsoon. \textit{Earth and Space Science}, \textbf{8~(6)}, e2020EA001\,471.

\bibitem[{Croitoru et~al.(2013)Croitoru, Piticar, Imbroane,, and Burada}]{croitoru2013}
Croitoru, A.-E., A.~Piticar, A.~M. Imbroane, and D.~C. Burada, 2013: Spatiotemporal distribution of aridity indices based on temperature and precipitation in the extra-carpathian regions of romania. \textit{Theoretical and applied climatology}, \textbf{112}, 597--607.

\bibitem[{Di et~al.(2017)Di, Duan, Gong, Ye,, and Miao}]{di2017parametric}
Di, Z., Q.~Duan, W.~Gong, A.~Ye, and C.~Miao, 2017: Parametric sensitivity analysis of precipitation and temperature based on multi-uncertainty quantification methods in the weather research and forecasting model. \textit{Science China Earth Sciences}, \textbf{60}, 876--898.

\bibitem[{Di et~al.(2018)Di, Duan, Wang, Ye, Miao,, and Gong}]{di2018assessing}
Di, Z., Q.~Duan, C.~Wang, A.~Ye, C.~Miao, and W.~Gong, 2018: Assessing the applicability of wrf optimal parameters under the different precipitation simulations in the greater beijing area. \textit{Climate dynamics}, \textbf{50}, 1927--1948.

\bibitem[{Di et~al.(2019)Di, Gong, Gan, Shen,, and Duan}]{Di2019}
Di, Z., W.~Gong, Y.~Gan, C.~Shen, and Q.~Duan, 2019: Combinatorial optimization for wrf physical parameterization schemes: A case study of three-day typhoon simulations over the northwest pacific ocean. \textit{Atmosphere}, \textbf{10}, \doi{10.3390/atmos10050233}.

\bibitem[{Di et~al.(2015)}]{di2015assessing}
Di, Z., and Coauthors, 2015: Assessing wrf model parameter sensitivity: A case study with 5 day summer precipitation forecasting in the greater beijing area. \textit{Geophysical Research Letters}, \textbf{42~(2)}, 579--587.

\bibitem[{Du~Duc et~al.(2019)Du~Duc, Hoang~Duc, Hole, Hoang, Luong Thi~Thanh,, and Mai~Khanh}]{du2019impacts}
Du~Duc, T., C.~Hoang~Duc, L.~R. Hole, L.~Hoang, H.~Luong Thi~Thanh, and H.~Mai~Khanh, 2019: Impacts of different physical parameterization configurations on widespread heavy rain forecast over the northern area of vietnam in wrf-arw model. \textit{Advances in Meteorology}, \textbf{2019~(1)}, 1010\,858.

\bibitem[{Evans et~al.(2012)Evans, Ekstr{\"o}m,, and Ji}]{evans2012evaluating}
Evans, J.~P., M.~Ekstr{\"o}m, and F.~Ji, 2012: Evaluating the performance of a wrf physics ensemble over south-east australia. \textit{Climate Dynamics}, \textbf{39}, 1241--1258.

\bibitem[{Evin et~al.(2014)Evin, Thyer, Kavetski, McInerney,, and Kuczera}]{evin2014comparison}
Evin, G., M.~Thyer, D.~Kavetski, D.~McInerney, and G.~Kuczera, 2014: Comparison of joint versus postprocessor approaches for hydrological uncertainty estimation accounting for error autocorrelation and heteroscedasticity. \textit{Water Resources Research}, \textbf{50~(3)}, 2350--2375.

\bibitem[{Gupta et~al.(2012)Gupta, Clark, Vrugt, Abramowitz,, and Ye}]{gupta2012towards}
Gupta, H.~V., M.~P. Clark, J.~A. Vrugt, G.~Abramowitz, and M.~Ye, 2012: Towards a comprehensive assessment of model structural adequacy. \textit{Water Resources Research}, \textbf{48~(8)}.

\bibitem[{Hastings(1970)}]{hastings1970monte}
Hastings, W.~K., 1970: Monte carlo sampling methods using markov chains and their applications. \textit{Biometrika}.

\bibitem[{Hung et~al.(2023)Hung, Tien, Quan, Duc, Dung, Hole,, and Nam}]{Hung2023}
Hung, M.~K., D.~D. Tien, D.~D. Quan, T.~A. Duc, P.~T.~P. Dung, L.~R. Hole, and H.~G. Nam, 2023: Assessments of use of blended radar–numerical weather prediction product in short-range warning of intense rainstorms in localized systems (swirls) for quantitative precipitation forecast of tropical cyclone landfall on vietnam’s coast. \textit{Atmosphere}, \textbf{14}, \doi{10.3390/atmos14081201}.

\bibitem[{Javanshiri et~al.(2021)Javanshiri, Fathi,, and Mohammadi}]{Javanshiri2021}
Javanshiri, Z., M.~Fathi, and S.~A. Mohammadi, 2021: Comparison of the bma and emos statistical methods for probabilistic quantitative precipitation forecasting. \textit{Meteorological Applications}, \textbf{28}, \doi{10.1002/met.1974}.

\bibitem[{Li et~al.(2020)Li, Ding,, and Li}]{li2020quantitative}
Li, X., R.~Ding, and J.~Li, 2020: Quantitative study of the relative effects of initial condition and model uncertainties on local predictability in a nonlinear dynamical system. \textit{Chaos, Solitons \& Fractals}, \textbf{139}, 110\,094.

\bibitem[{Liu et~al.(2021)Liu, Guo, Zhang, Zhang, Guan,, and Xu}]{Liu2021}
Liu, C., J.~Guo, B.~Zhang, H.~Zhang, P.~Guan, and R.~Xu, 2021: A reliability assessment of the ncep/fnl reanalysis data in depicting key meteorological factors on clean days and polluted days in beijing. \textit{Atmosphere}, \textbf{12}.

\bibitem[{Lorenz(1963)}]{lorenz1963}
Lorenz, E.~N., 1963: Deterministic nonperiodic flow. \textit{Journal of atmospheric sciences}, \textbf{20~(2)}, 130--141.

\bibitem[{Lorenz(1965)}]{lorenz1965}
Lorenz, E.~N., 1965: A study of the predictability of a 28-variable atmospheric model. \textit{Tellus}, \textbf{17~(3)}, 321--333.

\bibitem[{Lowrey and Yang(2008)Lowrey, and Yang}]{lowrey2008assessing}
Lowrey, M. R.~K., and Z.-L. Yang, 2008: Assessing the capability of a regional-scale weather model to simulate extreme precipitation patterns and flooding in central texas. \textit{Weather and Forecasting}, \textbf{23~(6)}, 1102--1126.

\bibitem[{Mai et~al.(2020)Mai, Craig,, and Tolson}]{mai2020simultaneously}
Mai, J., J.~R. Craig, and B.~A. Tolson, 2020: Simultaneously determining global sensitivities of model parameters and model structure. \textit{Hydrology and Earth System Sciences}, \textbf{24~(12)}, 5835--5858.

\bibitem[{McMillan et~al.(2012)McMillan, Krueger,, and Freer}]{mcmillan2012benchmarking}
McMillan, H., T.~Krueger, and J.~Freer, 2012: Benchmarking observational uncertainties for hydrology: rainfall, river discharge and water quality. \textit{Hydrological Processes}, \textbf{26~(26)}, 4078--4111.

\bibitem[{Moosavi et~al.(2021)Moosavi, Rao,, and Sandu}]{moosavi2021machine}
Moosavi, A., V.~Rao, and A.~Sandu, 2021: Machine learning based algorithms for uncertainty quantification in numerical weather prediction models. \textit{Journal of Computational Science}, \textbf{50}, 101\,295.

\bibitem[{Moosavi and Sandu(2018)Moosavi, and Sandu}]{moosavi2018}
Moosavi, A., and A.~Sandu, 2018: A state-space approach to analyze structural uncertainty in physical models. \textit{Metrologia}, \textbf{55~(1)}, S1.

\bibitem[{Morokoff and Caflisch(1995)Morokoff, and Caflisch}]{William1995}
Morokoff, W.~J., and R.~E. Caflisch, 1995: Quasi-monte carlo integration,. \textit{Journal of Computational Physics}, \textbf{122}, 218--230.

\bibitem[{Nasrollahi et~al.(2012)Nasrollahi, AghaKouchak, Li, Gao, Hsu,, and Sorooshian}]{nasrollahi2012assessing}
Nasrollahi, N., A.~AghaKouchak, J.~Li, X.~Gao, K.~Hsu, and S.~Sorooshian, 2012: Assessing the impacts of different wrf precipitation physics in hurricane simulations. \textit{Weather and Forecasting}, \textbf{27~(4)}, 1003--1016.

\bibitem[{{National Centers for Environmental Prediction, National Weather Service, NOAA, U.S. Department of Commerce}(2015)}]{cisl_rda_dsd083003}
{National Centers for Environmental Prediction, National Weather Service, NOAA, U.S. Department of Commerce}, 2015: Ncep gdas/fnl 0.25 degree global tropospheric analyses and forecast grids. Research Data Archive at the National Center for Atmospheric Research, Computational and Information Systems Laboratory, Boulder CO, \urlprefix\url{https://rda.ucar.edu/datasets/dsd083003/}.

\bibitem[{Rasmussen(2003)}]{rasmussen2003gaussian}
Rasmussen, C.~E., 2003: Gaussian processes in machine learning. \textit{Summer school on machine learning}, Springer, 63--71.

\bibitem[{Razavi et~al.(2012)Razavi, Tolson,, and Burn}]{razavi2012review}
Razavi, S., B.~A. Tolson, and D.~H. Burn, 2012: Review of surrogate modeling in water resources. \textit{Water Resources Research}, \textbf{48~(7)}.

\bibitem[{Sawada(2020)}]{sawada2020machine}
Sawada, Y., 2020: Machine learning accelerates parameter optimization and uncertainty assessment of a land surface model. \textit{Journal of Geophysical Research: Atmospheres}, \textbf{125~(20)}, e2020JD032\,688.

\bibitem[{Sawada and Duc(2024)Sawada, and Duc}]{sawada2024efficient}
Sawada, Y., and L.~Duc, 2024: An efficient and robust estimation of spatio-temporally distributed parameters in dynamic models by an ensemble kalman filter. \textit{Journal of Advances in Modeling Earth Systems}, \textbf{16~(2)}, e2023MS003\,821.

\bibitem[{Skamarock et~al.(2019)}]{skamarock2019description}
Skamarock, W.~C., and Coauthors, 2019: A description of the advanced research wrf model version 4 (vol. 145). \textit{National Center for Atmospheric Research}.

\bibitem[{Sloughter et~al.(2007)Sloughter, Raftery, Gneiting,, and Fraley}]{Sloughter2007}
Sloughter, J. M.~L., A.~E. Raftery, T.~Gneiting, and C.~Fraley, 2007: Probabilistic quantitative precipitation forecasting using bayesian model averaging. \textit{Monthly Weather Review}, \textbf{135}, 3209--3220, \doi{10.1175/MWR3441.1}.

\bibitem[{van~der Linden et~al.(2017)van~der Linden, Fink, Pinto,, and Phan-Van}]{vanderLinden2017}
van~der Linden, R., A.~H. Fink, J.~G. Pinto, and T.~Phan-Van, 2017: The dynamics of an extreme precipitation event in northeastern vietnam in 2015 and its predictability in the ecmwf ensemble prediction system. \textit{Weather and Forecasting}, \textbf{32}, 1041--1056, \doi{10.1175/WAF-D-16-0142.1}.

\bibitem[{Wang et~al.(2021)}]{wang2021quantifying}
Wang, C., and Coauthors, 2021: Quantifying physical parameterization uncertainties associated with land-atmosphere interactions in the wrf model over amazon. \textit{Atmospheric Research}, \textbf{262}, 105\,761.

\bibitem[{Wang et~al.(2023)Wang, Mo, Di, Liu, Lang,, and Duan}]{wang2023}
Wang, H., H.~Mo, Z.~Di, R.~Liu, Y.~Lang, and Q.~Duan, 2023: Knee point-based multiobjective optimization for the numerical weather prediction model in the greater beijing area. \textit{Geophysical Research Letters}, \textbf{50~(23)}, e2023GL104\,330.

\bibitem[{Wang and Tan(2023)Wang, and Tan}]{XuanWang2023}
Wang, X., and Z.~M. Tan, 2023: On the combination of physical parameterization schemes for tropical cyclone track and intensity forecasts in the context of uncertainty. \textit{Journal of Advances in Modeling Earth Systems}, \textbf{15}, \doi{10.1029/2022MS003381}.

\bibitem[{Yokoi and Matsumoto(2008)Yokoi, and Matsumoto}]{Yokoi2008}
Yokoi, S., and J.~Matsumoto, 2008: Collaborative effects of cold surge and tropical depression-type disturbance on heavy rainfall in central vietnam. \textit{Monthly Weather Review}, \textbf{136~(2)}, 3275--3287.

\bibitem[{Zhang et~al.(2006)Zhang, Odins,, and Nielsen-Gammon}]{zhang2006}
Zhang, F., A.~M. Odins, and J.~W. Nielsen-Gammon, 2006: Mesoscale predictability of an extreme warm-season precipitation event. \textit{Weather and forecasting}, \textbf{21~(2)}, 149--166.

\bibitem[{Zhu et~al.(2015)Zhu, Kong, Ran,, and Lei}]{Zhu2015}
Zhu, J., F.~Kong, L.~Ran, and H.~Lei, 2015: Bayesian model averaging with stratified sampling for probabilistic quantitative precipitation forecasting in northern china during summer 2010. \textit{Monthly Weather Review}, \textbf{143}, 3628--3641, \doi{10.1175/MWR-D-14-00301.1}.

\end{thebibliography}

\end{document}